\documentclass[aps,pre,reprint,superscriptaddress,showpacs]{revtex4-1}
\usepackage{graphicx}
\usepackage{dcolumn}
\usepackage{bm}
\usepackage{epsfig}

\input epsf
\epsfclipon

\begin{document}
\title{Information slows down hierarchy growth}

\author{Agnieszka Czaplicka}
\author{Krzysztof Suchecki}
\affiliation{Faculty of Physics, Center of Excellence for Complex Systems Research, Warsaw University of Technology, Koszykowa 75, PL-00-662 Warsaw, Poland}
\author{Borja Minano}
\author{Miquel Trias}
\affiliation{Institute for Applied Computing with Community Code $(\mathrm{IAC}^3)$ Universitat de les Illes Balears. Cra. Valldemossa km 7.5, E-07122, Palma, Spain}
\author{Janusz A. \surname{Ho\l yst}}
\affiliation{Faculty of Physics, Center of Excellence for Complex Systems Research, Warsaw University of Technology, Koszykowa 75, PL-00-662 Warsaw, Poland}
\affiliation{Netherlands Institute for Advanced Study in the Humanities and Social Sciences, Meijboomlaan 1, 2242 PR Wassenaar, The Netherlands}

\date{\today}
\begin{abstract}
We consider models of growing multi-level systems wherein the growth process is driven by rules of tournament selection.
A system can be conceived as an evolving tree with a new node being attached to a contestant node at the best hierarchy level (a level nearest to the tree root).
The proposed evolution reflects limited information on system properties available to new nodes.
It can also be expressed in terms of population dynamics.
Two models are considered: a constant tournament (CT) model wherein the number of tournament participants is constant throughout system evolution, and a proportional tournament (PT) model where this number increases proportionally to the growing size of the system itself.
The results of analytical calculations based on a rate equation fit well to numerical simulations for both models.
In the CT model all hierarchy levels emerge but the birth time of a consecutive hierarchy level increases exponentially or faster for each new level.
The number of nodes at the first hierarchy level grows logarithmically in time, while the size of the last, ``worst'' hierarchy level oscillates quasi log-periodically.
In the PT model the occupations of the first two hierarchy levels increase linearly but worse hierarchy levels either do not emerge at all or appear only by chance in early stage of system evolution to further stop growing at all.
The results allow to conclude that information available to each new node in tournament dynamics restrains the emergence of new hierarchy levels and that it is the absolute amount of information, not relative, which governs such behavior.
\end{abstract}
\maketitle

\section{Introduction}
In the area of complex systems, evolving networks have become intensively studied topic following the introduction of Barabasi-Albert (BA) model \cite{BA,jeong}, which offered a possible microscopic explanation of scale-free degree distribution emergence matching a broad range of real scale-free networks.
The tenet of the BA model, namely a preferential attachment rule, holds that a newly added node in a network connects to the already existing nodes with probabilities proportional to their current degrees.
Various forms of preference have been studied, e.g. nonlinear preferential attachment model of Krapivsky-Redner \cite{kr1,kr2}.
The preference may refer to diverse node attributes, such as their degree (\cite{BA,ravasz}), attractiveness \cite{attract}, fitness \cite{fit1,fit2} or age \cite{age}.
The above rules of preferential attachment require full information about the values of a relevant node attribute in entire network.
Such information is usually infeasible in respect to larger, real systems.
It makes highly relevant the question how the imposed information limit -- the limit in the amount of information new nodes possess about entire system -- impacts the growth process of a network.
In fact, a preferential attachment where information is local and limited has already been studied before \cite{xiang,aldridge,stefancic1}.
Therein new nodes could attach only to a random subset of all nodes.
A different approach than preferential attachment was considered in \cite{stefancic2}.
There, each new node was deterministically attached to a few highest degree nodes of a randomly chosen subset, thereby producing a scale-free topology over a range of degrees.
This procedure can be thought of as a ``tournament'' selection, wherein out of randomly chosen pool of participants only the best nodes ``win'' and get connected to.\\
While the preference in attachment may concern any given property of a node, our work focuses solely on a \textit{hierarchy level}.
This level corresponds to node's position within a certain hierarchical structure and may be relevant in social systems which are by nature ordered by some relations.
We have selected the hierarchy level to be an observable defining system dynamics both because hierarchies constitute a backbone of many complex systems and because node position in such a hierarchy often plays a decisive role.\\
In fact, the very concept of hierarchy has not yet been thoroughly explored and there exists no single, agreed upon definition of a ``hierarchy'' \cite{pumain}.
The concept of hierarchy has been applied in investigation of such diverse properties as importance of a node in a community structure \cite{copelli,vazquez}, participation of a node in activity patterns in neural networks \cite{yamashita}, a node's importance as a potential communication channel \cite{trusina}, or a node's relational importance in a knowledge structure \cite{muchnik}.
Lane \cite{pumain4} distinguishes the following kinds of hierarchy:
\begin{enumerate}
\item {\it order hierarchy}, where elements are ordered according to increasing or decreasing values of ordering variable, e.g. cities that are ordered according to their size \cite{pumain4,pumain6} or firms ordered according to their market capitalization,
\item {\it inclusion hierarchy} describing a nested structure of given entities, e.g. a holding consisting of companies, consisting of departments, consisting of offices etc. \cite{simon1,simon2},
\item {\it level hierarchy}, where entities are posited to some levels corresponding to scales/types of interactions and a set of interacting entities of a lower level comprises a higher level entity, e.g. biological ordering of cells comprising organ, of organs comprising individual, of individual comprising species \cite{simon1,simon2,anderson2,holland},
\item {\it control hierarchy} where elements are ordered according to direction of control, e.g. an officer of a higher rank can give an order to a lower rank officer or a Prime Minister can instruct a Minister who can further instruct a Department Director etc.
\end{enumerate}
While the order hierarchy has been studied since longer time \cite{bonabeau}, recent studies in complex networks area mainly address inclusion or level hierarchies \cite{copelli,vazquez,yamashita,trusina,muchnik}.
The inclusion hierarchy has been applied to synthetic models of hierarchical networks \cite{ravasz} and the level hierarchy has been studied in intracellular and intercellular networks, e.g. \cite{guimera,barabasi2}.\\
Here, we shall consider the control hierarchy that is typical for directed networks where arcs define ``higher-lower'' relations between nearest-neighbouring nodes, see e.g. \cite{murtra,mones}.
For simplicity we shall study a network in a form of a tree graph since it has a natural root definition and natural control relations between directly connected nodes.
Moreover, one can naturally define different hierarchy levels and thus such a system can be considered a ``perfect'' hierarchy \cite{murtra}.
Various dynamics have been investigated in such a topology \cite{sibani,cosenza,huberman,NSR} and the dynamics of the topology itself has also been well researched, including real systems such as internet news groups and forums \cite{kujawski} or directory trees \cite{klemm,geipel}.\\

Let us consider the growth of a tree graph, with new nodes being attached to the existing ones in respect to their hierarchy levels and a limited access to information about entire system.
We assume that new nodes representing social agents will try to occupy the best place in the existing social hierarchy.
Information constraints are modelled through limiting a random set of old nodes that they can connect to.
Thereby, a new node connects to an old one at the best possible level of hierarchy in the subset of known nodes.
While we limit ourselves to tree graphs, this is only a representation of a more general system that could also be considered in terms of population dynamics.\\
The aim of this study is to examine how the imposed information limit influences system structure, or more precisely, its influence on the emergence of consecutive hierarchy levels.
Our research is motivated by social dynamics where issues of limited information \cite{akcay,taborsky,barrett} have been recognised.
In fact, the amount of information available to community members has been considered in the perspective of evolving behavioral patterns \cite{taborsky} and the emergence of cooperation \cite{akcay}.
It has also been shown that individual information constraints can significantly alter the way in which cooperation arises \cite{akcay} and that high information costs lead to a steeper social hierarchy \cite{barrett}.
To the best of our knowledge, however, no studies have addressed the question how the amount of available information influences the growth of hierarchical networks which is the very aim of this work.\\

The paper is organized as follows. Section \ref{sec:2} introduces two models of tree evolution with a tournament selection where the number of contestants is constant (CT model) or it is proportional (PT model) to the current tree size. 
Sections \ref{sec:3} and \ref{sec:4} present analytical and numerical results of hierarchy growth in CT and PT models. Section \ref{sec:5} covers the main conclusions of our work.

\section{Tournament models} \label{sec:2}

\begin{figure}[ht]
 \centerline{ \epsfig{file=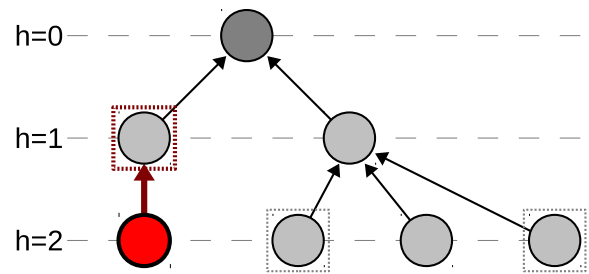,width=0.9\columnwidth} }
 \caption{(Color online) Evolving hierarchical system in the form of a tree, with marked hierarchy levels $h$. A new node (red) is informed of several randomly chosen existing nodes (marked with a square) and chooses the one at best position in the hierarchy (lowest $h$). It enters the hierarchy at a level worse by $1$ than a chosen node (in this example, at $h=2$). Note that new level $h=3$ emerges only when the random set of nodes contains only nodes positioned at level $h=2$.} \label{Fig:graf1} 
\end{figure}

Our model is a growing tree model, where at each time step we add a new node and choose one existing node to connect it to (see Fig. \ref{Fig:graf1}).
We take the tree to be as a hierarchical system, with hierarchy levels $h$ defined in respect to the distance from the tree root.
We call our system a hierarchy regardless of the actual emergent topology, including cases where it does not appear hierarchical at all.\\

The root node, a node at level $h=0$ (the ``top'' level) is created at time step $t=0$.
There is always only one root, since introducing new roots would require adopting additional dynamical rules, while a single initial root can be considered the initial condition.
Level $h=0$ will be considered \emph{the best} hierarchy level, while the following $h=1$, $h=2$ etc. will be considered \emph{worse} levels.
To avoid confusion, we refer to these levels as \emph{better/worse} or \emph{older/younger} instead of \emph{higher/lower}, since a smaller $h$ value is better and can be considered ``higher'' in the hierarchy.
At each time step $t$ a new node is added (which means that the system size is $N=t+1$), and then it is connected to one of the existing nodes $j$ at a hierarchy level $h_j$.
A new node $i$ is therefore at hierarchy level $h_i=h_j+1$ and this does not change in time.\\
The model could be also understood in terms of population dynamics, with populations $N_h$ of individuals occupying different hierarchy levels $h$.
At each time step $t$ a new agent $i$ enters the system, with one of existing agents $j$ becoming his ``superior''.
The hierarchy $h_i$ of newcomer is one step worse than his immediate superior's, meaning $h_i=h_j+1$.
This approach is equivalent to the tree representation, provided that rules of nodes attachment are dependent only on node hierarchies $h_j$, and not on other nodes properties, e.g. nodes degrees.\\

The choice of a node to connect to is that of a ``tournament selection'' \cite{stefancic2}, where a subset of $m$ random nodes is selected from among the nodes already present, and a node at the best hierarchy level (lowest $h$) from among these is chosen to be connected to.
Limited size of a tournament reflects limited availability of information \cite{informacja}, whereby the choice is limited to the set of known nodes.
Two tournament variants will be studied: a model with a given \emph{constant tournament} size $m$ (CT model) and a model with random, \emph{proportional tournament} size (PT model).\\
In the \emph{CT model}, the number of contestants $m>0$ is a fixed integer parameter and thus new nodes have access to constant amount of information.
The contestants are randomly chosen without repetition from among the existing nodes.
For $t<m-1$ all the existing nodes are participants.
When $m=1$, the tournament has only one contestant and the tree is randomly growing.\\
In the \emph{PT model} each existing node has a fixed probability $\alpha$ of participating in the tournament.
This means that the number of participants is a random variable, with an average $\langle m \rangle (t)=\alpha \cdot t$ dependent on time.
For higher $t$ the distribution of tournament sizes $m$ is Poissonian.
This means that the average amount of information available to new nodes increases with time.
In case the contest ends up with no participants, it is redone until at least one contestant is present.
Such cases occur mainly in the initial stage of the tree evolution when $\langle m \rangle (t) \leq 1$.
The re-doing of empty tournament events ensures that it is possible for a new node to attach somewhere at every time step.\\
Although our approach to network dynamics takes into account a selection pressure, it considerably differs from both the Barabasi-Albert model of evolving networks \cite{BA} and all similar models (e.g.\cite{kr1},\cite{attract}) using preferential attachment where a temporary node degree defines the probability of selection.
In our models the level of hierarchy of the existing nodes does not change when new nodes attach to it, unlike the degree of nodes in BA model.
Moreover, the mechanism of selection takes into account limited amount of information available to the nodes being attached.

\section{Emergence of hierarchy levels in constant tournament model} \label{sec:3}
\subsection{General approach}

Since the edges do not play any role in the considered attachment dynamics, one can ignore the details of evolving network topology and analyze only the numbers of nodes $N_h$ at each hierarchy level $h$.
We base our analytical description on the average outcomes of the processes and use rate equations to describe the dynamics of averages.
Because new roots do not emerge, the number of nodes at level $h=0$ is constant in time and equal to $N_0(t)=1$.
Occupations $N_h(t)$ of every level $h>0$ can grow in time when a node is added to level $h$.
It takes place in a situation when the best hierarchy level in the tournament is $h-1$, meaning that the set of randomly chosen nodes at time $t$ includes at least one node from level $h-1$ and $m-1$ nodes from levels $h \geq h-1$.
It follows that the rate equation can be written in the continuous time approximation as

\begin{equation}
\frac{dN_h(t)}{dt}=\frac{ \left( \begin{array}{c} N_{h-1}^+(t) \\ \\ m \end{array} \right) - \left( \begin{array}{c} N_h^+(t) \\ \\ m \end{array} \right) }{ \left( \begin{array}{c} N(t) \\ \\ m \end{array} \right) } \text{   for } h>0 \label{rate}
\end{equation}

where we denote $N_h^+$ as the number of nodes at hierarchy level $h$ and worse
\begin{equation}
N_h^+ = \sum_{i=h}^{+\infty} N_i = N - \sum_{i=0}^{h-1} N_i \label{eq_nhplus}
\end{equation}
The following initial conditions will be used for Eq. (\ref{rate}):  
\begin{equation}
 N_0(0)=1, \text{ and } N_{h>0}(0)=0 \label{init} 
\end{equation}

Since the complexity makes the rate equation (\ref{rate}) generally unsolvable, the following Sections will consider its specific cases.

\subsection{Number of nodes at different hierarchy levels}
\textit{For $m=1$} the CT model has only one contestant and the dynamics is reduced to a random selection process.

\begin{figure}[ht]
\centerline{\psfig{file=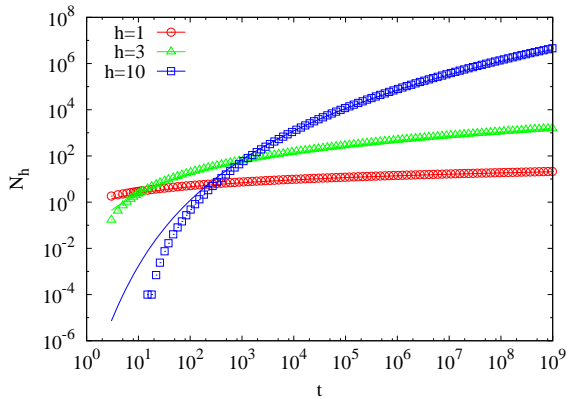,width=0.9\columnwidth}}
\caption{(Color online) \textit{CT model}. Number of nodes $N_h(t)$ at each hierarchy level $h$ increases as a logarithm to power $h$ when the attachment process is random ($m=1$). Results of computer simulations are presented by symbols, analytical results are presented by solid lines of corresponding colors and follow from Eq. (\ref{Nhm1t}). Note that a younger hierarchy level grows faster than older ones.}
\label{Fig:Nhm1}
\end{figure}

The rate equation (\ref{rate}) simplifies to :
\begin{equation}
\frac{dN_h(t)}{dt}=\frac{N_{h-1}(t)}{N(t)}, \text{ for } h >0 \label{Eq:Nhm1}
\end{equation}
which corresponds to the probability that a randomly chosen node will be at level $h-1$.
The solution of Eq. (\ref{Eq:Nhm1}) with initial conditions (\ref{init}) can be written as:
\begin{equation}
N_h(t)= \frac{1}{h!} \left(\ln\left(t+1\right)\right)^h \label{Nhm1t}
\end{equation}
Numerical and analytical results are presented in Fig. \ref{Fig:Nhm1}. Results of computer simulations at this and at all following plots have been averaged over $Q=10^4$ realisations.
It is evident that except for small times $t$, our analytic approach correctly captures the dynamics of $N_h$.
Eq. (\ref{Nhm1t}), normalized by $N=t+1$ takes the form of a Poisson distribution of nodes at hierarchy levels $h$, with the mean value increasing as a logarithm of time $\langle h \rangle \sim \ln (t+1)$.
This kind of log-poissonian statistics appears in the dynamics of various complex systems dominated by short events (called ``quakes''), separated by increasingly long times of inactivity \cite{anderson}.
The log-poissonian distribution of the number of these events after the lapse of time $t$ arises from the constant probability of event happening in logarithmic time scale $P(t_1,t_2) \sim \ln(t_2)-\ln(t_1)$.
While such quake distribution arises from aggregation of different possible realizations, Eq. (\ref{Nhm1t}) represents a distribution over different nodes in a single network.
Although both distributions are alike, no direct relation between dynamics of both systems could be established.\\

\begin{figure}[ht]
 \centerline{ \epsfig{file=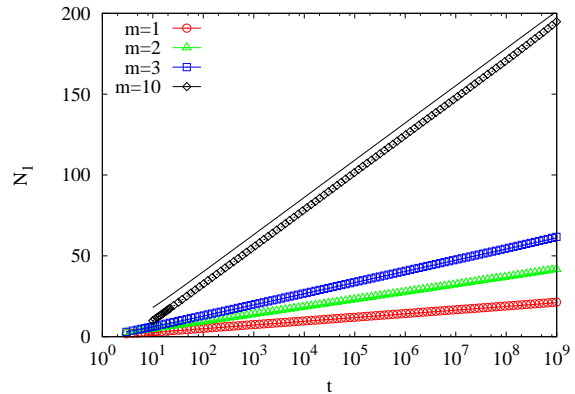,width=0.9\columnwidth} }
 \caption{(Color online) \textit{CT model}. Logarithmic growth of number of nodes $N_1(t)$ at the first hierarchy level $h=1$ for different sizes of the tournament $m=1,2,3,10$. Results of computer simulations are presented by symbols, analytical results from Eq. (\ref{eqh1}) are presented by solid lines of corresponding colors.} \label{Fig:N1m}
\end{figure}

\textit{For $m>1$} and $h=1$ the Eq. (\ref{rate}) can be also simplified, realizing that for $h=1$ we have $N_{h-1}^+=N(t)$ and $N_h^+(t)=N(t)-1$. Thus
\begin{equation}
\frac{dN_1(t)}{dt}=\frac{m}{t+1}
\end{equation}
This means that the number of nodes at the first hierarchy level increases logarithmically with $t$ and after taking into account the initial condition (\ref{init}) one gets:
\begin{equation}
N_1(t)=m\ln(t+1)
\label{eqh1}
\end{equation}
which is in agreement with the numerical data (see Fig. \ref{Fig:N1m}).

\begin{figure}[ht]
 \centerline{ \epsfig{file=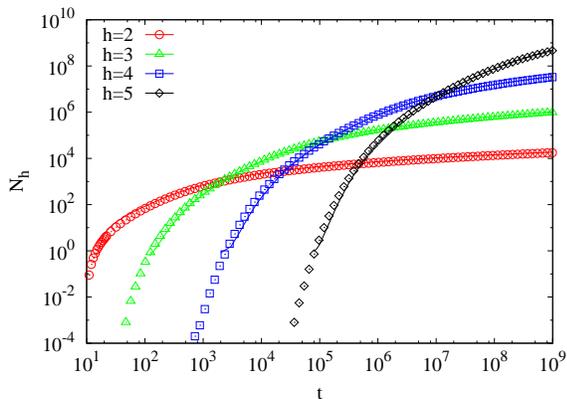,width=0.9\columnwidth} }
 \caption{(Color online) \textit{CT model}. Hierarchy levels that were born earlier grow slower than the ones following them. The graph shows the evolution of hierarchy level occupancy $N_h(t)$ for $m=10$. Computer simulations are presented by symbols,  numerical solutions of the rate Eq. (\ref{rate}) are presented by solid lines of corresponding colors.} \label{Fig:Nhm10}
\end{figure}

What happens at the following levels?
For $h>1$ we have not found analytical formula for $N_h(t)$, but the numerical integration of the rate equation  Eq. (\ref{rate}) displays a good agreement with numerical simulations of the tournament process  (see Fig. \ref{Fig:Nhm10}).
It is worth noting that for times $t$ smaller than $\tau_h$ moment of emergence of hierarchy level $h$ (see next Section) the variable $N_h(t)$ received from the integration of gamma function appearing in Eq. (\ref{rate}) can have non-physical values (i.e. negative or imaginary).
For that reason we have performed a numerical integration of Eq. (\ref{rate}) starting from $t=\tau_h$ and taking into account the initial conditions $N_h(t=\tau_h)=1$.
Values $\tau_h$ for  $h=1,2\ldots$ have been found in simulations of the tree growth (see Section \ref{sec:tauh}).
Fig. \ref{Fig:Nhm10} demonstrates   that $N_h(t)$ reveals an interesting and nontrivial behavior.
Each following hierarchy level $h$ grows faster than the previous one (levels that were born earlier than $h$) and consequentially after some time the number of nodes at level $h$ is greater than the number of nodes at other levels.
A similar behavior is observed in Fig. \ref{Fig:Nhm1}.
In fact, if in the a given time step the hierarchy level $h$ makes for the largest number of nodes $N_h(t)$ then the majority of new nodes are attached to nodes at this same level, and the following level $h+1$ grows very fast and it eventually becomes the most occupied one.

\begin{figure}[ht]
 \centerline{ \epsfig{file=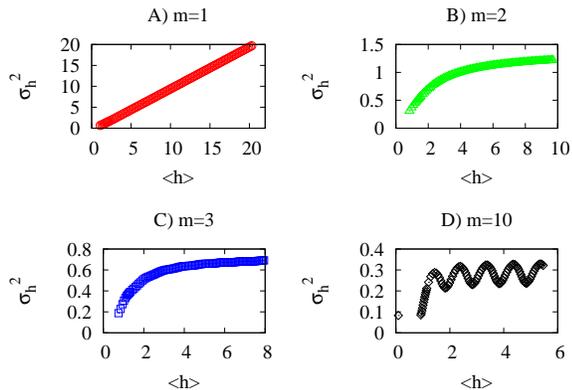,width=0.9\columnwidth} }
 \caption{(Color online) \textit{CT model}. The variance $\sigma^2$ of the distribution of levels $h$ does not scale linearly with the mean $\langle h \rangle$ for the dynamics for $m>1$, which implies non-poissonian character. The oscillations are the result of discrete $h$ values and are minimal when $\langle h \rangle$ is close to an integer number. Data obtained from numerical simulations averaged over $Q=10^4$ realisations.} \label{Fig:sigmaCT}
\end{figure}

The poissonian character of the distribution of levels $h$ in the system (Eq.(\ref{Nhm1t})), evident for $m=1$ does not extend to cases when $m>1$.
In fact, numerical simulations show, that the variance of $h$ distribution does not scale linearly with mean $\langle h \rangle$ for $m>1$ (Fig. \ref{Fig:sigmaCT}).

\subsection{Hierarchy level birth time} \label{sec:tauh}

In terms of mean values, the emergence of hierarchy level $h$ at time $\tau_h$ means that the expected number of nodes at this level equals one:
\begin{equation}
N_h(\tau_h)=1
\end{equation}
For \textit{CT model} taking the solution (\ref{Nhm1t}) one can write the hierarchy level birthtimes as
\begin{equation}
\tau_h=\exp\left[\left(h!\right)^{\frac{1}{h}}\right]-1 \label{tauhm1eq}
\end{equation}

\begin{figure}[ht]
\centerline{\psfig{file=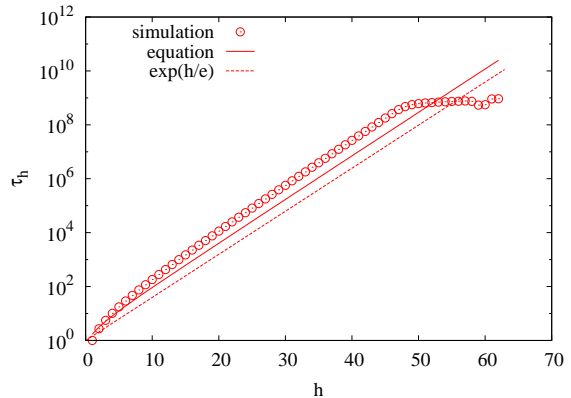,width=0.9\columnwidth}}
\caption{(Color online) \textit{CT model}. For $m=1$ new hierarchy levels emerge exponentially with time. Results of computer simulations are presented by circles,  analytical results following from Eq. (\ref{tauhm1eq}) by a solid line. The plateau for higher $h$ is a consequence of a limited simulation length ($10^9$ time steps). Dashed line shows the exponential  behavior $\tau_h \propto \exp(h/e)$ }
\label{Fig:tauhm1} 
\end{figure}

The solution (\ref{tauhm1eq}) and numerical simulations are presented in Fig. \ref{Fig:tauhm1}. Let us note that 
\begin{equation}
\lim_{h\rightarrow \infty}\frac{\tau_{h+1}}{\tau_h}=e^{1/e}\approx 1.44  
\end{equation}
thus for large $h$ the birth time of new levels increases exponentially, $\tau_h \sim \exp(h/e)$

\begin{figure}[ht]
\centerline{\psfig{file=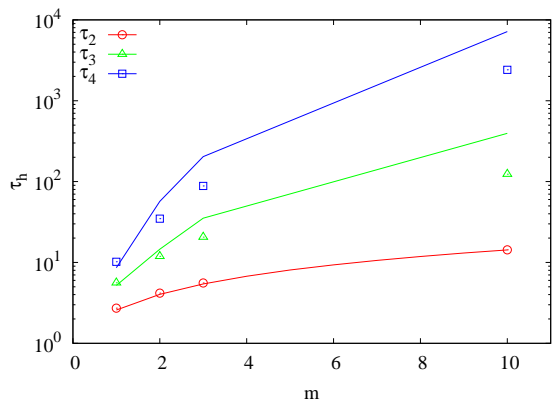,width=0.9\columnwidth}}
\caption{(Color online) \textit{CT model}. Birth time $\tau_h$ for a given hierarchy level $h$ is delayed when size $m$ of the tournament increases. Results of computer simulations are presented by symbols.  Solid line for  $\tau_2(m)$ follows from  Eq. (\ref{eqtau2m}) and  solid lines for $\tau_3(m)$ and $\tau_4(m)$ from  Eq. (\ref{tauhmnum}).}
\label{Fig:tau2m}
\end{figure}

The mean time $\tau_2$ can be analytically calculated by considering the exact process (not the averages).
We start with one node at hierarchy level $h=0$.
At time steps $t\leq m$ we add nodes at level $h=1$.
Starting from time step $m$, there is a chance to create hierarchy level $h=2$ -- prior to that all nodes are always created at level $h=1$.
Since  $m$ nodes for the tournament are chosen out of the total of $t+1$ nodes, there is ${{t+1}\choose{m}}$ such combinations in total.
Out of all of them there is ${{t}\choose{m}}$ combinations where the root is not chosen, giving the probability of creating hierarchy level $h=2$ at time step $t+1$ (provided it was not created before):
\begin{equation}
P_2(t+1)=\frac{{{t}\choose{m}}}{{{t+1}\choose{m}}} = 1-\frac{m}{t+1}
\end{equation}
Since the probability to continue the growth of level $h=1$ is 
\begin{equation}
P_1(t+1)=1-P_2(t+1)=\frac{m}{t+1}
\end{equation}
the total probability that the process will create the hierarchy level $h=2$ exactly at time step $t$ is
\begin{equation}
P(t) = \left[ \prod_{k=m+1}^{t-1} P_1(k) \right] P_2(t) = m^{t-m-1} \frac{m!}{(t-1)!}(1-\frac{m}{t})
\end{equation}
It follows that the mean time $\tau_2$ when the hierarchy level $h=2$ appears is
\begin{equation}
\tau_2 = \sum_{t=m+1}^{+\infty} P(t) t = m \left[ 1 + \left( \frac{\mathrm{e}}{m} \right)^m \left( \Gamma(m) - \Gamma(m,m) \right) \right] \label{eqtau2m}
\end{equation}
where $\Gamma(m)$ is Euler's gamma function and $\Gamma(m,m)$ is an incomplete gamma function.
In case $m=1$ this formula can be greatly simplified and one gets a value close to the result (\ref{tauhm1eq})
\begin{equation}
\tau_2|_{m=1}= \mathrm{e} \label{eq_tau2m1}
\end{equation}
The solution (\ref{eqtau2m}) is in a very good agreement with the numerical simulations presented in Fig. \ref{Fig:tau2m}.

To find $\tau_h$ for $h>2$ and $m>1$ one should take into account that the emergence of new hierarchy levels is inhibited by the presence of nodes at all older, better levels.
Thus, it is more difficult to precisely determine the time when a new level will emerge for the first time.
A new level $h$ can appear if level $h-1$ consists of at least $m$ nodes, but it is very unlikely that such an event will take place at time $\tau_{h-1} + m$.
Therefore, one should take into account not only the number of nodes at $N_{h-1}$ but also each number $N_i$ for $i<h$.
Without analytical solutions for $N_h(t)$ when $m>1$ and $h>1$ we are unable to find an analytical result for $\tau_h(h)$ in the way presented above.
Instead of that, we can estimate time $\tau_h$ as follows.
Between times $\tau_{h-1}$ and $\tau_h$ we added $N(\tau_h)-N(\tau_{h-1})$ nodes divided between all existing hierarchy levels:

\begin{equation}
N(\tau_h)-N(\tau_{h-1})=\sum_{i=0}^h{\left(N_i(\tau_h)-N_i(\tau_{h-1})\right)}
\end{equation}\label{deltatauhm}

When a new level $h$ emerges there is $N_h(\tau_h)=1$, and $N_h(t<\tau_h)=0$.
Taking into account these two assumptions and knowing that $N(t)=t+1$ we obtain:

\begin{equation}
\tau_h - \sum_{i=0}^{h-1}{N_i(\tau_h)}=\tau_{h-1}- \sum_{i=0}^{h-2}{N_i(\tau_{h-1})} \label{tauhmnum}
\end{equation}

\begin{figure}[ht]
\centerline{\psfig{file=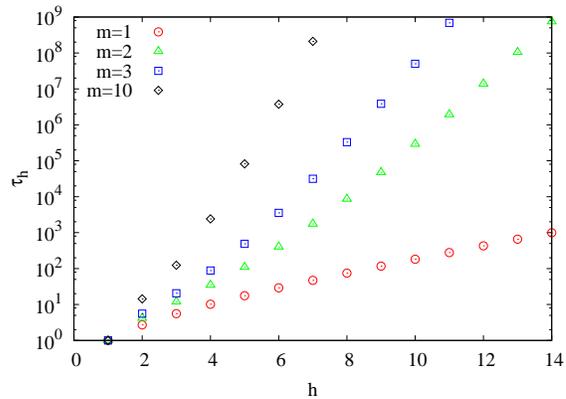,width=0.9\columnwidth}}
\caption{(Color online) \textit{CT model}. For $m>1$ the rate of hierarchy growth is slower than the exponential behavior observed in case $m=1$. Results of computer simulations for hierarchy level birth times $\tau_h$ are presented by different symbols for different tournament sizes $m$. Larger $m$ slows down the emergence of new hierarchy levels.}
\label{Fig:tauhm}
\end{figure}

Since we know $\tau_2(m)$ and have numerical results for $N_h(t)$ we can find numerical solution of Eq. (\ref{tauhmnum}) for a given $h>2$ (see Fig. \ref{Fig:tau2m}).
Fig. \ref{Fig:tauhm} presents times of birth $\tau_h$  for various hierarchy levels $h$ and for various tournament sizes $m$.
The emergence of the following hierarchy levels is slower for higher $m$.
It means that the more nodes are participating in a tournament, the lower the chance of a new level to appear.
If $m\to \infty$ a new node is equipped with full information about all node hierarchy levels and thus all nodes attach to the hub, i.e. the network becomes a star graph and hierarchy levels $h>1$ do not emerge.

\subsection{Number of nodes at maximal hierarchy level $h_{max}$}
\begin{figure}[ht]
\centerline{\psfig{file=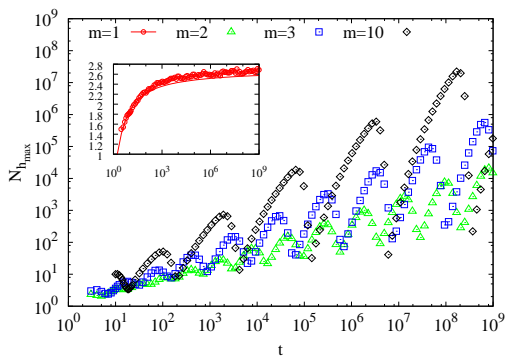,width=0.8\columnwidth}}
\caption{(Color online) \textit{CT model}. For $m>1$ number of nodes $N_{h_{max}}(t)$ at maximal hierarchy level oscillates log-periodically with time. For $m=1$ the number $N_{h_{max}}(t)$ increases slowly and oscillations are negligible. The solid line in the inset is an estimation of $N_{h_{max}}(t)$ from combinations of  Eqs. (\ref{Nhm1t}), (\ref{tauhm1eq}).}
\label{Fig:Nmaxm}
\end{figure}

At each time step $t$ one can distinguish the maximal hierarchy level (the worst one) $h=h_{max}$ in the network structure.
Let us consider the changes of a number of nodes $N_{h_{max}}(t)$ in time at such a level.\\
\textit{For the case $m=1$} corresponding to the random attachment process one can use the explicit solutions (\ref{Nhm1t}), (\ref{tauhm1eq}) and after some algebra one receives that $N_{h_{max}}$ oscillates between $1$ and $N^{max}_{h_{max}}=\exp\left[\left(h_{max}+1\right)!^{1/\left(h_{max}+1\right)}\right]$.
The last value corresponds to time $\tau_{h_{max}+1}$ when hierarchy level $h_{max}$ is replaced by hierarchy $h_{max}+1$ being the worst one.
Let us note that for $h_{max} \rightarrow +\infty$ the value $N^{max}_{h_{max}}$ reaches the limit $N^{max}_{\infty}=\mathrm{e}$, and thus the amplitudes of such oscillations are small. 
Since value $\tau_{h_{max}+1}$ is given by Eq. (\ref{tauhm1eq}) one gets the estimate of $N_{h_{max}}$ in a discrete set of time steps.
In Fig. \ref{Fig:Nmaxm} inset the result is compared to numerical simulations.
Since the oscillations of $N_{h_{max}}$ are small, they are invisible when averaged over many realizations of tree dynamics.         

It can be understood as follows.
When we consider only one contestant, a new hierarchy level $h$ can emerge if level $h-1$ contains at least one node.
For this reason the level considered to be the maximal level $h_{max}$ at one time step can in principle give birth to a new level at the following time step when a new node is attached to it.
Of course, when network is large it takes longer for such a node to be selected.\\
\textit{For $m > 1$} the number of nodes at the maximal level $h_{max}$ depends on stochastically determined hierarchy level birth times $\tau_h$ and is nonmonotonous.
The numerical dependence $N_{h_{max}}(t)$ is shown in Fig. \ref{Fig:Nmaxm}.
In this case hierarchy level $h$ can emerge if level $h-1$ is present in the system and the number of nodes at this level is large enough ($N_{h-1}(\tau_h)\geq m$).
It follows that level $h-1$ must attract a larger number of nodes than for $m=1$.
The actual $N_{h_{max}}$ increases until a new hierarchy level is born at $\tau_h$, which makes $N_{h_{max}}=1$.
Consecutively, number $N_{h_{max}}=1$ begins to grow again, forming a quasi-log-periodic pattern.
The average over many different realizations transforms such a pattern into the observed log-periodic oscillations of $N_{h_{max}}$, with amplitude increasing along tournament size $m$ and time $t$.

\subsection{Total number of hierarchy levels}

\begin{figure}[ht]
 \centerline{ \epsfig{file=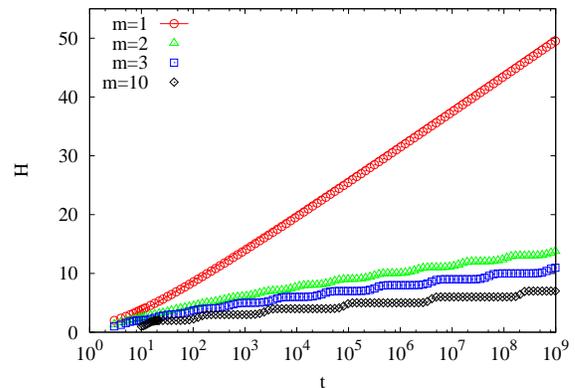,width=0.9\columnwidth} }
 \caption{(Color online) \textit{CT model}. Logarithmic growth of the number of hierarchy levels in time $H(t)$ for $m=1$ and step-like growth for $m>1$. Results of computer simulations are presented by points, analytical results for $m=1$ (Eq. (\ref{eqHm1})) by a solid line.}
 \label{Fig:Hm}
\end{figure}

Let us consider how on average the total number of hierarchy levels $H(t)$ increases in time.
Since we start labeling levels from $h=0$ thus $H(t)= h_{max}+1$.\\
\textit{For $m=1$} following hierarchy levels $h$ emerge at time steps $\tau_h$ which are given by Eq. (\ref{tauhm1eq}).
Using Eq. (\ref{tauhm1eq}) and Stirling formula $n!=\left(\frac{n}{e}\right)^n\sqrt{2\pi n}$ we can obtain an approximate solution:

\begin{equation}
H(t) \approx e\ln(t+1)\left[2\pi e\ln\left(t+1\right)\right]^{-\frac{1}{e\ln\left(t+1\right)}} \label{eqHm1}
\end{equation}

In the limit of large $t$ there is  
\begin{equation} H_{large}(t) \approx e\ln(t+1) \label{approxHm1}
\end{equation}  
The solution (\ref{eqHm1}) fits well to the numerical simulations presented in Fig. \ref{Fig:Hm}. 

\textit{For $m > 1$} we could not find an analytical formula for $H(t)$  since  we do not have analytical form of $\tau_h$.
Fig. \ref{Fig:Hm} shows numerical simulations of this observable.
The behavior of $H(t)$ for $m>1$ is more complex than in the case of random growth $m=1$.
We observe a step-like growth being the consequence of discrete values of hierarchy level $h$ separated by much longer time-spans between emergence of new levels at times $\tau_h$ than for $m=1$.
The larger the tournament $m$ the slower the increase of the total number of hierarchy levels $H(t)$ and the more evident the step-like behavior of such a process. 

\subsection{Mean hierarchy approach} \label{analityka}
We can approximate the hierarchy evolution in time by calculating the mean hierarchy of newly added nodes.
Since the nodes are added one per a time step, we can identify a node by time step $\tau$ it was added to the network.
The mean hierarchy level of a node added at time $t$ is equal to the mean level of the node it attaches to plus $1$.
Using the continuous variable approximation, the above can be written as
\begin{equation}
h(t) = 1 + \int \limits_{0}^{t} P(\tau,t) h(\tau) \mathrm{d}\tau \label{eq_mean_continuous}
\end{equation}
where $P(\tau, t)$ is a probability to attach to a node $\tau$ at time step $t$.
The probability $P(\tau, t)$ depends on the parameter $m$.
For $m=1$ it is simply a chance to randomly pick one node out of $t$ existing nodes
\begin{equation}
P(\tau,t)=\frac{1}{t}
\end{equation}
If $m>1$, then we need to take into account that not all of the nodes have the same probability of being chosen.
We simplify our approach by assuming that an older node will be on average at a better hierarchy level than a node added at a later time step.
The hierarchy level relation is therefore by assumption mapped on the age relation. \\
Since there is $m$ chances to pick a node of age $\tau$ for a tournament, the probability of such an event is $m/t$.
For the node to be the winner, it needs to be the oldest.
It means that all other $m-1$ nodes must be younger than $\tau$.
Thus the probability to have a node of age $\tau$ as the winner of the tournament is
\begin{equation}
P(\tau,t)=\frac{m}{t} \left( 1 - \frac{\tau}{t} \right) ^ {m-1}
\end{equation}
With this Eq. (\ref{eq_mean_continuous}) changes to
\begin{equation}
h(t)= 1 + \frac{m}{t} \int\limits_{0}^{t} \left( 1 - \frac{\tau}{t}\right)^{m-1} h(\tau) \mathrm{d}\tau \label{eq_mean_general}
\end{equation}
This equation can be solved by multiplying it by $t^m$ and differentiating $m$ times over $t$ sidewise, thereby transforming it into a differential equation.
The only solution not diminishing quickly to zero with time is
\begin{equation}
h(t) = a(m) \ln t + 1 \label{eq_assume_ht}
\end{equation}
with constant $1$ resulting from the imposed initial condition $h(1)=1$.
Putting it into Eq. (\ref{eq_mean_general}) allows one to find the value of factor $a(m)$ as
\begin{equation}
a(m) = \frac{1}{H_m} \label{eq_am}
\end{equation}
where $H_m$ is harmonic number $H_m=\sum_{i=1}^{m} 1/i$.

\begin{figure}[ht]
 \centerline{\epsfig{file=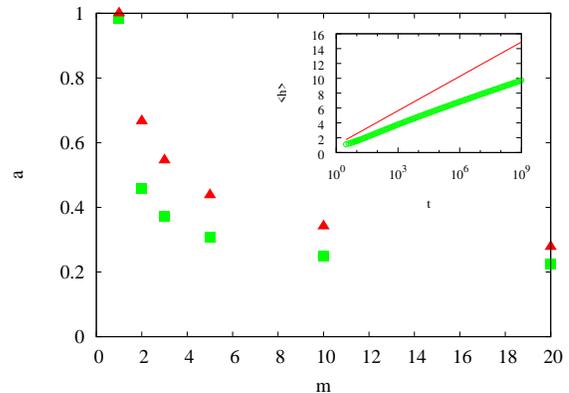,width=0.9\columnwidth}}
 \caption{(Color online) \textit{CT model}. Coefficient $a(m)$ corresponding to the rate of hierarchy level increase. Red triangles are results of computer simulations and green squares follow from Eq. (\ref{eq_am}). Inset: Evolution of average hierarchy level $\left\langle h\right\rangle (t)$ for $m=2$, analytical prediction (Eq.(\ref{eq_assume_ht}), line) and numerical results (symbols) showing different coefficients.} \label{Fig:avhm}
\end{figure}

Because the approach is based on mean values thus $h(t)$ behaves smoothly with time, but real $h(t)$ will be rather noisy.
To obtain a relatively smooth variable for comparison we consider the average hierarchy level in the network

\begin{eqnarray}
\left\langle h \right\rangle (t) = \frac{1}{t} \int\limits_0^t h(\tau) \mathrm{d}\tau = a \ln t + (1-a)
\label{alfameq}
\end{eqnarray}
Numerical results for $\langle h \rangle$ confirm (Fig. \ref{Fig:avhm}) that it grows logarithmically, as expected from Eq. (\ref{alfameq}).
The coefficients $a(m)$ were calculated by fitting logarithmic curves $h(t)=a(m) \ln t + \mathrm{const.}$ to the data.
Fig. \ref{Fig:avhm} shows comparison between numerical and analytical values of $a(m)$.
The exact values do not match, although the observed coefficients decrease with the tournament size $m$ and therefore with the availability of information, in accordance with analytical prediction (Eq. \ref{eq_am}).
We can conclude that analytical approach is successful in predicting the logarithmic behavior, but -- due to approximations we have used, it may not predict the exact values.
In fact, if we modify the model to prefer older, not best-level nodes (mirroring the approximation made in our analytical approach), then numerical results fit the analytic predictions, which means that the discrepancy comes from the ``older nodes have better hierarchy levels'' approximation.

\section{Emergence of hierarchy levels in proportional tournament model} \label{sec:4}

\subsection{Growth of hierarchy levels}
In the PT model, with proportional tournament size, all nodes have a fixed probability $\alpha$ to participate in a given tournament.
The overall tournament size is therefore a random variable, and in the limit of large time it has Poissonian distribution.
The mean tournament size $\langle m \rangle = \alpha t$ increases in time, thus on average the amount of information available for new nodes increases in time as well.
We will note the chance that a node does not participate in a tournament as $q=1-\alpha$.
Let us look at a single tournament, and how much on average the number of nodes at each hierarchy level $h$ changes afterwards.
Since the root level $h=0$ has always a single node $N_0=1$, therefore the mean change on this level is zero $\Delta N_0 = 0$ similarly as in the CT model. 
The level $h=1$ grows when the tournament is won by the root, which happens every time it participates, i.e. with probability $\alpha$.
Similar line of thought leads to a general rate equation for the mean change $\Delta N_h$ during a single tournament for any $h>1$
\begin{equation}
\Delta N_h = \left( 1-q^{N_{h-1}}\right) \cdot \prod_{i=0}^{h-2} q^{N_i}. \label{eq_mp_deltanh}
\end{equation}
In our model, if no nodes participate in the tournament, then the procedure is repeated until at least one node is present and can be the winner.
If the tournament ends up empty, then no node is added and the system time clock $t$ is at halt.
Since the chance there are no participants is $q^N=q^{t+1}$, we can write the mean change of the time clock during a single tournament as
\begin{equation}
\Delta t = 1-q^{t+1} \label{eq_mp_deltat}
\end{equation}
To simplify calculations, we use a continuous variables approach.
For $h=1$, Eqs. (\ref{eq_mp_deltanh}) and (\ref{eq_mp_deltat}) give the rate equation 
\begin{equation}
\frac{\mathrm{d} N_1}{\mathrm{d} t} = \frac{1-q}{1-q^{t+1}} \label{eq_mp_dn1}
\end{equation}
When $t \gg 1$ the solution is
\begin{equation}
N_1(t) \approx \alpha t \label{eq_mp_an1}
\end{equation}
thus hierarchy level $h=1$ grows linearly in the large time limit.
The rate equation for hierarchy levels $h \ge 2$ is
\begin{equation}
\frac{\mathrm{d} N_h}{\mathrm{d} t} = \frac{\left( 1-q^{N_{h-1}}\right) \cdot \prod_{i=0}^{h-2} q^{N_i}}{1-q^{t+1}} \label{eq_mp_dnh}
\end{equation}

\begin{figure}[ht]
 \centerline{\epsfig{file=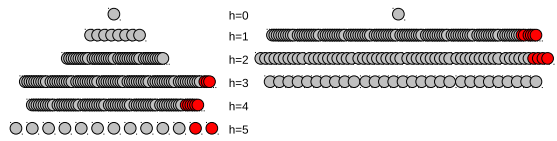,width=0.99\columnwidth}}
 \caption{(Color online) Examples of typical shapes of systems comprising $200$ nodes in the CT (left, $m=2$) and PT (right, $\alpha=0.5$) models. Only nodes at different levels are shown, the links are omitted for clarity. $10$ newest nodes are marked in red, thus showing at which levels the tree grows. Note that while in the CT model it is the worst levels that grow fast, allowing the emergence of new levels, in the PT model it is only levels $1$ and $2$ that do grow after lapse of time.} \label{Fig:przyklady}
\end{figure}

It follows that for large times the level $h=2$ also grows linearly 
\begin{equation}
N_2(t) \approx q t \label{eq_mp_an2}
\end{equation}
Thus for $t \gg 1$ we have $N_1(N)+N_2(N) \approx N$ which implies that new nodes only appear at levels $h=1$ and $h=2$, while worse levels do not grow at all.
This effect can be confirmed by looking at rate equation for levels $h>2$ and observing that their growth is limited by the presence of nodes at levels $h=1$ and $h=2$ in the form of terms $q^{N_1}$ and $q^{N_2}$.
Since the numbers $N_1$ and $N_2$ grow linearly, these terms go towards zero.
This behavior differs much from the evolution of CT model, where all hierarchy levels emerge and grow when given enough time (Fig. \ref{Fig:przyklady}).\\

\begin{figure}[ht]
 \centerline{\psfig{file=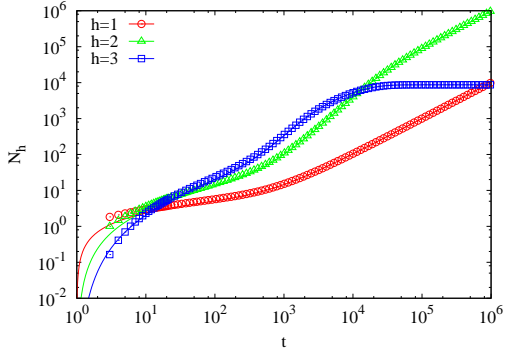,width=0.8\columnwidth}}
 \caption{(Color online) \textit{PT model}. The growth of large trees is monopolized by hierarchy levels $h=1$ and $h=2$, since in the course of time levels $h=3$ and worse stop growing. The graph shows $N_{h}(t)$ for $\alpha=0.01$. Results of computer simulations are presented by symbols,  corresponding analytical results (Eq. (\ref{eq_mp_n1}), (\ref{eq_mp_n2}))  by solid lines.  The number $N_3(t)$ shown as a numerical solution of Eq. (\ref{eq_mp_dnh}) saturates around time $t=10^4$.}
 \label{Fig:Nhmt}
\end{figure}

Let us look at system dynamics before the first two levels monopolize its growth.
The exact solution of Eq. (\ref{eq_mp_dn1}) is
\begin{equation}
N_1(t)=(1-q) t + (1-q) \frac{\ln (1-q)}{\ln q} - (1-q) \frac{\ln (1-q^{t+1})}{\ln q} \label{eq_mp_n1} 
\end{equation}
Similary Eq. (\ref{eq_mp_dnh}) can be solved for $h=2$, obtaining
\begin{eqnarray}
N_2(t)= q t + q \frac{\ln (1-q)}{\ln q} - q \frac{\ln (1-q^{t+1})}{\ln q} - \label{eq_mp_n2} \\
-\frac{1}{\ln q} \left( \frac{q}{1-q} \right)^q \left( \left( \frac{1-q^{t+1}}{q^{t+1}} \right)^{-(1-q)} - \left( \frac{1-q}{q} \right)^{-(1-q)} \right) \nonumber
\end{eqnarray}
For $h>2$ we could not find an analytic solution for the Eq. (\ref{eq_mp_dnh}) but we integrated it numerically to get $N_3(t)$ used in comparison with simulation results (Fig. \ref{Fig:Nhmt}).\\

\begin{figure}[ht]
 \centerline{ \epsfig{file=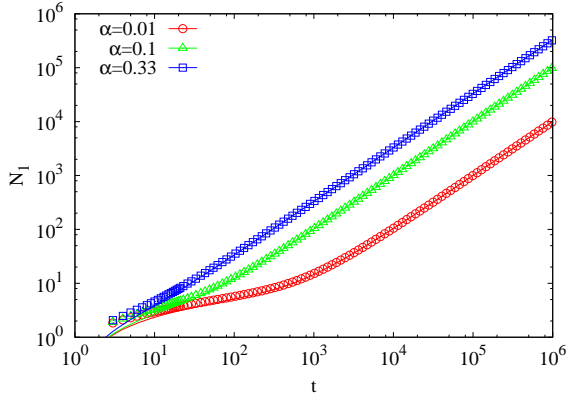,width=0.9\columnwidth} }
 \caption{(Color online) \textit{PT model}. The hierarchy level directly below the root ($h=1$) grows linearly with time. The graph shows $N_1(t)$ for $\alpha=0.01,0.1,0.33$. Results of computer simulations  are presented by symbols, analytical results (Eq. (\ref{eq_mp_n1})) by solid lines.}
 \label{Fig:N1mt}
\end{figure}

The Eq. (\ref{eq_mp_n1}) and (\ref{eq_mp_n2}) have linear behavior for $t \gg 1/\alpha$, but behave nonlinearly in the beginning.
Figs. \ref{Fig:N1mt} and \ref{Fig:Nhmt} show that the analytical predictions agree very well with the results obtained through numerical simulations of the model.
$N_1$ starts at the same point, and then evolve differently for different $\alpha$, only converging to appropriate $\alpha t$ line after some time, the longer the lower is $\alpha$.
$N_2$ grows slow at the beginning, but then accelerates as $\alpha t$ grow and finally converges to $q t$.
$N_3$ initially grows, but then it stops and creates a plateau, at about the same time $N_1$ and $N_2$ converge to their limit forms.

\begin{figure}[ht]
 \centerline{\epsfig{file=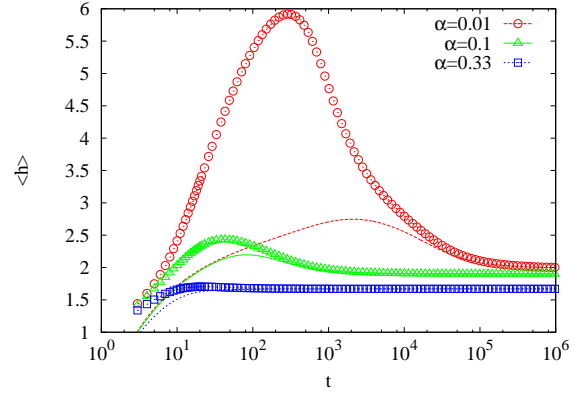,width=0.9\columnwidth}}
 \caption{(Color online) \textit{PT model}. After the initial increase of the number of hierarchy levels, the trend reverses and levels $1$ and $2$ monopolize growth, causing the average hierarchy level $\left\langle h\right\rangle (t)$ to be between $1$ and $2$. The graph shows the evolution of $\left\langle h\right\rangle (t)$ for different $\alpha=0.01,0.1,0.33$. Results of computer simulations are presented by symbols, approximate analytical solutions (Eq. (\ref{eq_mp_avh})) by lines.} \label{Fig:avhmt}
\end{figure}

Knowing that for larger times levels $h=1$ and $h=2$ contain almost all nodes of the tree, it is possible to determine the average hierarchy level in the graph at $N \to +\infty$.
It is simply
\begin{equation}
\langle h \rangle(t \to +\infty) \to \frac{N_1(t) + 2 N_2(t)}{t+1} \to 2 - \alpha \label{avhmteq}
\end{equation}
and therefore it is always between $1$ and $2$.
The shape of the evolution of average hierarchy level $\langle h \rangle(t)$ can be estimated (simply ignoring all hierarchies worse than $h=3$) as
\begin{equation}
\langle h \rangle(t) \approx \frac{N_1(t)+2 N_2(t)+3 N_3(t)}{N_1(t)+N_2(t)+N_3(t)} \label{eq_mp_avh}
\end{equation}
Fig. \ref{Fig:avhmt} shows the values obtained from numeric simulations as well as from Eqs. (\ref{eq_mp_n1}), (\ref{eq_mp_n2}) and numerical solution of Eq. (\ref{eq_mp_dnh}) for $h=3$.
The prediction attains the same general shape, increasing at the beginning and then falling down towards the limit value $2-\alpha$, although due to considering only first three levels one can not predict the full height of the peak, which is caused by the presence of nodes at many different levels, including those worse than $h=3$.

\subsection{Critical parameters for hierarchy level emergence}
\begin{figure}[ht]
 \centerline{\epsfig{file=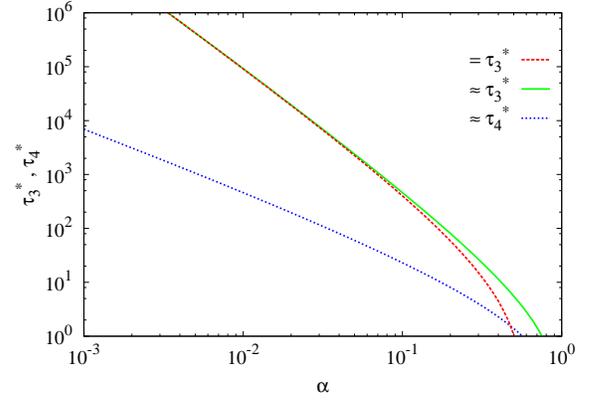,width=0.9\columnwidth}}
 \caption{(Color online) \textit{PT model}. The critical time $\tau^*_3$ of halting the growth of hierarchy levels $h=3$ or worse decreases with the probability $\alpha$ of a node selection for a single tournament. Worse levels stop growing even earlier and $\tau^*_4(\alpha) <\tau^*_3(\alpha)$. For $\alpha \ll 1$ the two approximations for $\tau^*_3$  represented by Eq. (\ref{eq_mp_taug1}) (red broken) and Eq. (\ref{eq_mp_taug2}) (green solid) are indistinguishable and time $\tau^*_4$ approximated by the solution (\ref{eq_mp_taug3}) (blue dotted) fullfills the condition  $\tau^*_4 \ll \tau^*_3$. \label{Fig:taug}}
\end{figure}

Similar to the case of CT model, the hierarchy level emergence times $\tau_h$ are very hard to describe analytically, aside from trivial $\tau_0=0$ and $\tau_1=1$.
For very small $\alpha$ if we take into account that we force at least one participant to be in tournament, the PT model can be approximated by the CT model with $m=1$.
This basically means random attachment, where $\tau_2 \approx \mathrm{e}$ (Eq. (\ref{eq_tau2m1})).
Since for higher $\alpha$ the chance to choose nodes at hierarchy level $h=1$ is lower than this approximation, the resulting time $\tau_2$ is actually higher.
For $\alpha \to 1$, the level $h=2$ never appears and $\tau_2 \to +\infty$.
Unlike in the CT model, where new hierarchy levels appear and grow, in the PT model, the first two monopolize the growth and appearance of levels $h=3$ or worse is not certain.
It is possible to calculate the time when the growth of all levels worse than $h=2$ stops.
First we write the equation for growth of hierarchy levels $h=3$ and worse
\begin{equation}
\frac{\mathrm{d} N_3^+}{\mathrm{d} t} = \frac{q^{N_1+1} \left( 1-q^{t-N_1-1}\right)}{1-q^{t+1}} \label{eq_mp_dn3plusdt}
\end{equation}
If we integrate it from $\tau$ to infinity, we obtain the number $N_3^{++}(\tau)$ of nodes of levels $h=3$ and worse that are \emph{expected} to appear after time $\tau$.
The exact expression is
\begin{equation}
N_3^{++} (\tau) = \frac{ -\left( \frac{q}{1-q} \right)^q}{\ln q} \cdot \left( \frac{q^{\tau+1}}{1-q^{\tau+1}} \right)^{(1-q)} - \frac{\ln(1-q^{\tau+1})}{\ln q} \label{eq_mp_n3p}
\end{equation}
Let us define $\tau^*_h$ as the critical time after which no nodes are expected to appear at levels $h$ or worse, and the growth is completely monopolized by better levels.
The condition for $\tau^*_3$ can be written as
\begin{equation}
N_3^{++} (\tau^*_3) = 1 \label{eq_mp_condtaug}
\end{equation}
Using Eqs. (\ref{eq_mp_n3p}) and (\ref{eq_mp_condtaug}) we get 
\begin{equation}
\tau^*_3 \approx \frac{\ln(-\ln q)}{(1-q) \ln q} + \frac{q}{1-q} \left( \frac{\ln (1-q)}{\ln q} - 1 \right) \label{eq_mp_taug1}
\end{equation}
Since the time $\tau^*_3$ decays with the probaility $\alpha$ (see Fig. \ref{Fig:taug}) thus one can calculate the critical $\alpha_3^*$, which is the maximum $\alpha$ where the levels worse than $2$ are expected to appear at all.
Putting $\tau^*_3=0$ and $N_3^{++}=1$ into Eq. (\ref{eq_mp_n3p}) we obtain equation for $\alpha_3^*$
\begin{equation}
\frac{1-\alpha_3^*}{\alpha_3^*} + \ln \alpha_3^* = - \ln (1-\alpha_3^*)
\end{equation}
It follows the critical value $\alpha_3^*\approx 0.4138$.\\

When $\alpha$ is small and $\tau^*_3$ is large, the critical value $\tau^*_3(\alpha)$ can also be estimated in a simpler way, without using Eq. (\ref{eq_mp_n3p}).
In such a case we approximate the growth of hierarchy levels $h=3$ and worse by ignoring the influence of empty tournaments and simplifying Eq. (\ref{eq_mp_dn3plusdt}) to
\begin{equation}
\frac{\Delta N_3^+}{\Delta t} \approx q^{N_1} \approx q^{\alpha t} \label{eq_mp_deltan3plus}
\end{equation}
The condition for critical $\tau^*_3$ can be then written as
\begin{equation}
\sum_{t=\tau^*_3}^{+\infty} q^{\alpha t} = 1
\end{equation}
which leads to the approximate solution
\begin{equation}
\tau^*_3 \approx \frac{-\ln(\alpha^2)}{\alpha^2} \label{eq_mp_taug2}
\end{equation}
Fig. \ref{Fig:taug} shows that both methods of approximation give  the same dependence of $\tau^*_3(\alpha)$ for small $\alpha$ values.
It also shows an estimate for the $\tau^*_4$ -- the time after which no nodes are expected to appear at levels $h=4$ or worse.
It could be obtained in the same way as the approximation (\ref{eq_mp_taug2}) for $\tau^*_3$, except instead of considering only level $h=1$ obstructing the growth through term $q^{N_1}$, we take into account both $q^{N_1}$ and $q^{N_2}$, which means we get
\begin{equation}
\frac{\Delta N_4^+}{\Delta t} \approx q^{N_1} q^{N_2} \approx q^t \label{eq_mp_deltan4plus}
\end{equation}
thus
\begin{equation}
\tau^*_4 \approx \frac{-\ln \alpha}{\alpha} \label{eq_mp_taug3}
\end{equation}

Note that in this approximation all hierarchy levels worse than $h=3$ behave the same way and one would obtain the same approximate for the critical time $\tau^*_h$ for any $h>3$, which is smaller than $\tau^*_3$.

\begin{figure}[ht]
 \centerline{ \epsfig{file=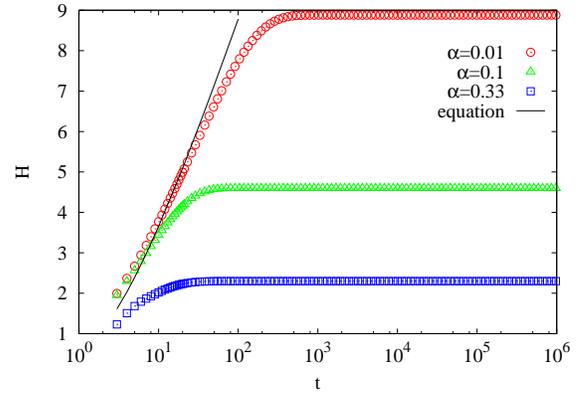,width=0.9\columnwidth} }
 \caption{(Color online) \textit{PT model}. Number of hierarchy levels $H(t)$ grows during initial time, and then saturates as first two levels monopolize the growth. The graph shows $H(t)$ for different values of $\alpha=0.01,0.1,0.33$. Results of computer simulations  are shown as symbols, analytical estimate (Eq. (\ref{eqHm1})) by the solid line.}
 \label{Fig:Hmt}
 \end{figure}

The number of hierarchy levels behaves as one would expect from earlier findings (Fig. \ref{Fig:Hmt}).
In the beginning, when $\alpha t<1$ it is increasing logarithmicaly and then it saturates.
This can be explained as follows.
When $t<1/\alpha$, the tournament has usually just one contestant.
In fact a probability of a larger tournament $m>1$ is $\alpha^{(m-1)}$ and it is small when  $\alpha \ll 1$. 
This means that the behavior is approximately the same as for the CT model, with $m=1$, and the maximum hierarchy level will grow in a similar fashion (Eq. (\ref{eqHm1}) ). 
Once $t>1/\alpha$, the growth of levels slows down and eventually stops, as the size of the tournament $\langle m \rangle$ increases (Fig. \ref{Fig:Hmt}).\\
The number of nodes $N_{h_{max}}(t) $ at the maximum hierarchy level $h_{max}$ evolves in time as follows.
When $\alpha>\alpha_3^*$ the level $h=3$ does not appear at all, and $h_{max}=2$, thus $N_{h_{max}}$ grows linearly with time.
When $\alpha<\alpha_3^*$  levels worse than $2$ do appear and then stop growing, meaning that $N_{h_{max}}(t)$ is constant from that time on.\\

\section{Conclusions} \label{sec:5}
We conclude that in a tree growth where nodes attach to the best known place in hierarchy the availability of information restrains the emergence of hierarchy levels -- the larger the amount of available information the slower the growth of consecutive hierarchies.
The non-trivial observation is that it is the absolute amount of information, not relative, that governs this behavior.
If new nodes know about a constant number of existing nodes, then the system grows steadily, as in the CT model (Fig. \ref{Fig:Hm}).
If new nodes know about a fixed fraction of existing nodes, then the system dynamics changes in time and hierarchy growth slows down to a complete standstill, as in the PT model (Fig. \ref{Fig:Hmt}).
This is because information about only one well positioned node is required for the new node to connect well, regardless of how many nodes there are in total.
Repeated connections to nodes at good hierarchy levels make it even easier for new nodes to connect well, producing very wide and shallow tree (Fig. \ref{Fig:przyklady}).
This behavior resembles models of group cooperation, where easy access to information causes a hierarchy to become shallower \cite{barrett} provided that system resources are evenly distributed.
Considering that the CT and PT models differ only in respect to the dependency of information on system size and yet display qualitatively different behavior, we may conclude that there must exist a transition between these two types of behavior and, consequently, a critical dependence of information on system size.
What is the actual critical dependence for stopping a hirerachy growth is yet an open question.

\begin{acknowledgments}
The research leading to these results has received funding from the European Union Seventh Framework Programme (FP7/2007-2013) under grant agreement no 317534 (the Sophocles project) and from the Polish Ministry of Science and Higher Education grant 2746/7.PR/2013/2.
J.A.H. has been also supported by a Grant from The Netherlands Institute for Advanced Study in the Humanities and Social Sciences (NIAS) and by European Union COST TD1210 KNOWeSCAPE action.
\end{acknowledgments}


\begin{thebibliography}{10}
\bibitem{informacja} The information measure could be formally defined as mean logarithm of the tournament size i.e. $I=\left\langle \log(m) \right\rangle$. For CT model it is $I_{CT}=\log(m)$, while for PT model it is more complicated, only converging to $I_{PT}\approx \log(\alpha t)$ for $t\to +\infty$. Random attachment ($m=1$) corresponds to $I_{CT}(1)=0$.
\end{thebibliography}
\end{document}